\documentclass[10pt,twocolumn]{article}
\usepackage{makeidx}
\usepackage[french, english]{babel}
\usepackage[latin1]{inputenc}
\usepackage{amsfonts}
\usepackage{bm}
\usepackage{amssymb}
\usepackage{latexsym}
\usepackage{amsmath}
\usepackage{amscd}
\usepackage{amsthm}
\usepackage{hyperref}
\usepackage{rotating}
\usepackage{graphicx}
\usepackage{pifont}
\usepackage{ulem}
\usepackage{curves}
\usepackage{fancyhdr}
\usepackage{epsfig}
\usepackage{pstricks}
\usepackage{subfigure}
\usepackage{epic}
\usepackage{alltt}
\usepackage[super,sort&compress,comma]{natbib}

\def\GA{\mathbf{GA}}
\def\CRL{\mathbf{CRL}}
\def\FA{\mathbf{FA}}
\theoremstyle{plain}

\title{New first trimester crown-rump length's equations optimized by structured data collection from a French general population}

\author{Marc Constant\thanks{corresponding author: Marc@Constant.eu ; Centre de radiologie et d'imagerie médicale, 386 avenue de Dunkerque, 59130 Lambersart, France},\hspace{0.5cm} Viet Chi Tran\thanks{Laboratoire Paul Painlevé UMR CNRS 8524, UFR de Maths, Université des Sciences et Technologies Lille 1, Cité Scientifique, 59655 Villeneuve d'Ascq Cedex, France},\hspace{0.5cm} Bernard Benoît\thanks{Centre Femme Mère Enfant, Hôpital de l'Archet 2, Nice, France},\hspace{0.5cm} Francis Vasseur\thanks{Biostatistics department EA2694, University Hospital, University Lille Nord de France, Pôle de Santé Publique Parc Eurasanté CHRU, 59037 Lille Cedex, France}}

\begin{document}

\maketitle

\begin{abstract}
\textbf{Objectives} Prior to foetal karyotyping, the likelihood of Down's syndrome is often determined combining maternal age, serum free beta-HCG, PAPP-A levels and embryonic measurements of crown-rump length and nuchal translucency for gestational ages between 11 and 13 weeks. It appeared important to get a precise knowledge of these scan parameters' normal values during the first trimester. This paper focused on crown-rump length.\\

\noindent \textbf{Methods} 402 pregnancies from in-vitro fertilization allowing a precise estimation of foetal ages (FA) were used to determine the best model that describes crown-rump length (CRL) as a function of FA. Scan measures by a single operator from 3846 spontaneous pregnancies representative of the general population from Northern France were used to build a mathematical model linking FA and CRL in a context as close as possible to normal scan screening used in Down's syndrome likelihood determination. We modeled both CRL as a function of FA and FA as a function of CRL. For this, we used a clear methodology and performed regressions with heteroskedastic corrections and robust regressions. The results were compared by cross-validation to retain the equations with the best predictive power. We also studied the errors between observed and predicted values.\\

\noindent \textbf{Results} Data from 513 spontaneous pregnancies allowed to model CRL as a function of age of foetal age. The best model was a polynomial of degree 2. Datation with our equation that models spontaneous pregnancies from a general population was in quite agreement with objective datations obtained from 402 IVF pregnancies and thus support the validity of our model. The most precise measure of CRL was when the SD was minimal (1.83mm), for a CRL of 23.6 mm where our model predicted a 49.4 days of foetal age. Our study allowed to model the SD from 30 to 90 days of foetal age and offers the opportunity of using Zscores in the future to detect growth abnormalities.\\

\noindent \textbf{Conclusion} With powerful statistical tools we report a good modeling of the first trimester embryonic growth in the general population allowing a better knowledge of the date of fertilization useful in the ultrasound screening of Down's syndrome. The optimal period to measure CRL and predict foetal age was 49.4 days (9 weeks of gestational age). Our results open the way to the detection of foetal growth abnormalities using CRL Zscores throughout the first trimester.\\

\noindent \textbf{Keywords:} crown-rump length; foetal age; equation of the standard deviation; normal parameters; structured embryonic data collection.
\end{abstract}

\section{Introduction}
In France, prior to foetal karyotyping, the likelihood of chromosomal aneuploidy is determined with an algorithm that combines maternal age, serum free $\beta$-HCG, pregnancy-associated plasma protein A levels and scan embryonic measurements of first trimester crown-rump length (CRL) and nuchal translucency \cite{cuckle}\cite{cucklewaldthompson}\cite{herman}. When the probability of aneuploidy exceeds 1/250, foetal karyotyping is proposed, whatever the woman's age. The window of screening requires scan measures for gestational age (GA) between 11 weeks + 0 day and 13 weeks + 6 days (CRL between 45 and 84 mm)\cite{arrete}, thus during and at the boundary of the first trimester. To date the most consensual reference regarding CRL according to the foetal age (FA) (FA=GA-14), arose from Robinson in 1973 \cite{robinson73} on an Australian population. Regardless of their validity, Robinson's data were obtained using ultrasound technologies far from the modern ultrasound devices now routinely used and given the great potential impact of ultrasound measures on the medical decision in the aneuploidy diagnosis process, it appeared crucial to get a precise knowledge of CRL variations with current ultrasound technologies. Our study of the first trimester CRL variation in a French general population, completes the results of Salomon et al.\cite{salomon06} regarding the second and third trimesters.\\
Since Robinson, several CRL curves have been proposed. Hadlock\cite{hadlock} published an equation that integrates data from vaginal measurements, Verwoerd-Dikkeboom and al.\cite{verwoerd} proposed an equation obtained with measures stemming from a three-dimensional acquisition. In 2008 Verbrug et al.\cite{verburg} and in 2010 Papaioannou et al.\cite{papaioannou} established new charts based on multiethnic samples taken by multiple operators with various devices. In 2010, Pexters and al. \cite{pexsters} established a multi-operator curve on a population sample close to that in our study. Our curves did not significantly differed from the classical Robinson curve, and those recently presented\cite{verwoerd}\cite{papaioannou}\cite{pexsters}. The present work provides a first trimester external validation of these recent charts and pleads for the use of our equations in routine foetal aneuploidy screening at least in the French population. Moreover, our work offers for the first time a modeling of the standard deviation, which is a prerequisite step towards the definition of normal foetal growth parameters.

\section{Methods}

\subsection{Data}

402 pregnancies from in-vitro fertilization (IVF) were ultrasound measured between February 1989 and September 2010 by a single operator (B.B.) in Nice and Monaco scan centers. In this sample, non-viable embryos were excluded. There were 78 (19.4\%) twin and 8 (1.99\%) triplet pregnancies. In case of a multiple pregnancy, only one embryo measurement was retained. Comparison of CRL between single and multiple pregnancies was performed with the Wilcoxon and Chow tests. Wilcoxon signed-rank test compares the two samples and tests whether their population means differ \cite{sheskin}. This was possible since the FA distributions were similar between groups. However, since the CRL clearly depends on FA, the Chow test was more suitable as it compares the modeling of CRL as a function of FA and checks whether the same regression can be applied to both samples or whether two different models are needed \cite{dougherty}\cite{huber}\cite{maronna}\cite{sheskin}.\\

The general population sample consisted of 3846 spontaneous pregnancies examined in one center (Lambersart) by a single operator (M.C.) between January 2007 and February 2011. CRLs were measured for 2123 of them. Pregnancies from IVF were excluded. To determine as accurately as possible the date of fecundation, we included in the study only pregnancies of women who had regular periods with a cycle duration that was clearly expressed, and who knew the date of their last menstruations and who had no contraception for at least 3 months before pregnancy. As it is well known that CRL measurements are less accurate on large embryos, only CRL measures below 85 mm were kept. Thus 513 pregnancies fitted our selection criteria and were scored. As estrus occurs 14 days before menstruations rather than at half of the menstrual cycle (due to delayed ovulation\cite{lynch}), the J0 of life for the foetus was determined as 14 days before the next estimated menstruation. In case of a multiple pregnancy, the measurement of only one embryo was kept.\\

\begin{figure*}[!ht]
     \centering
\begin{tabular}{ccc}
(a) & (b) & (c)\\
 \hspace{-0.5cm}    \includegraphics[width=6cm]{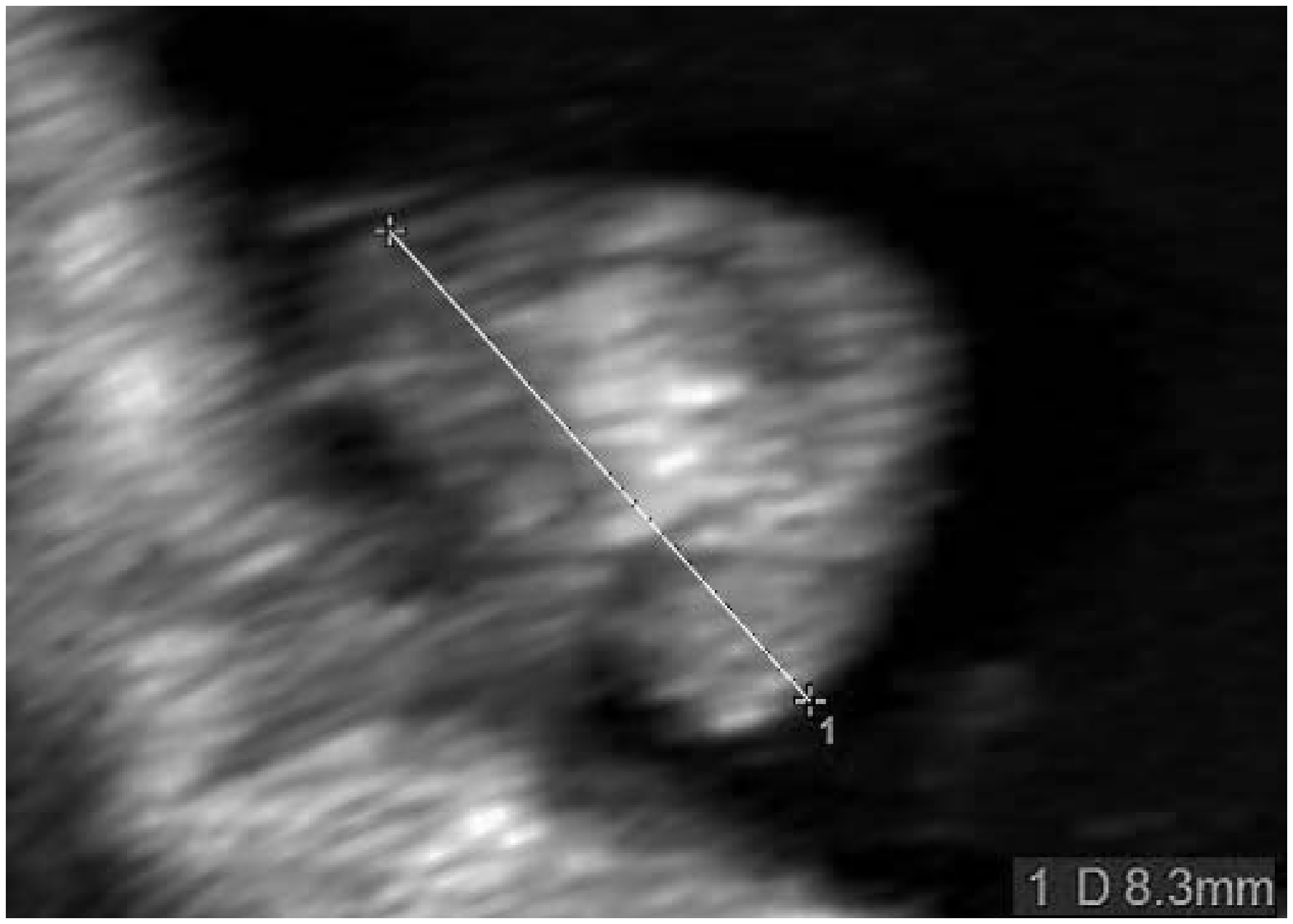}
&  \hspace{-0.5cm}   \includegraphics[width=6cm]{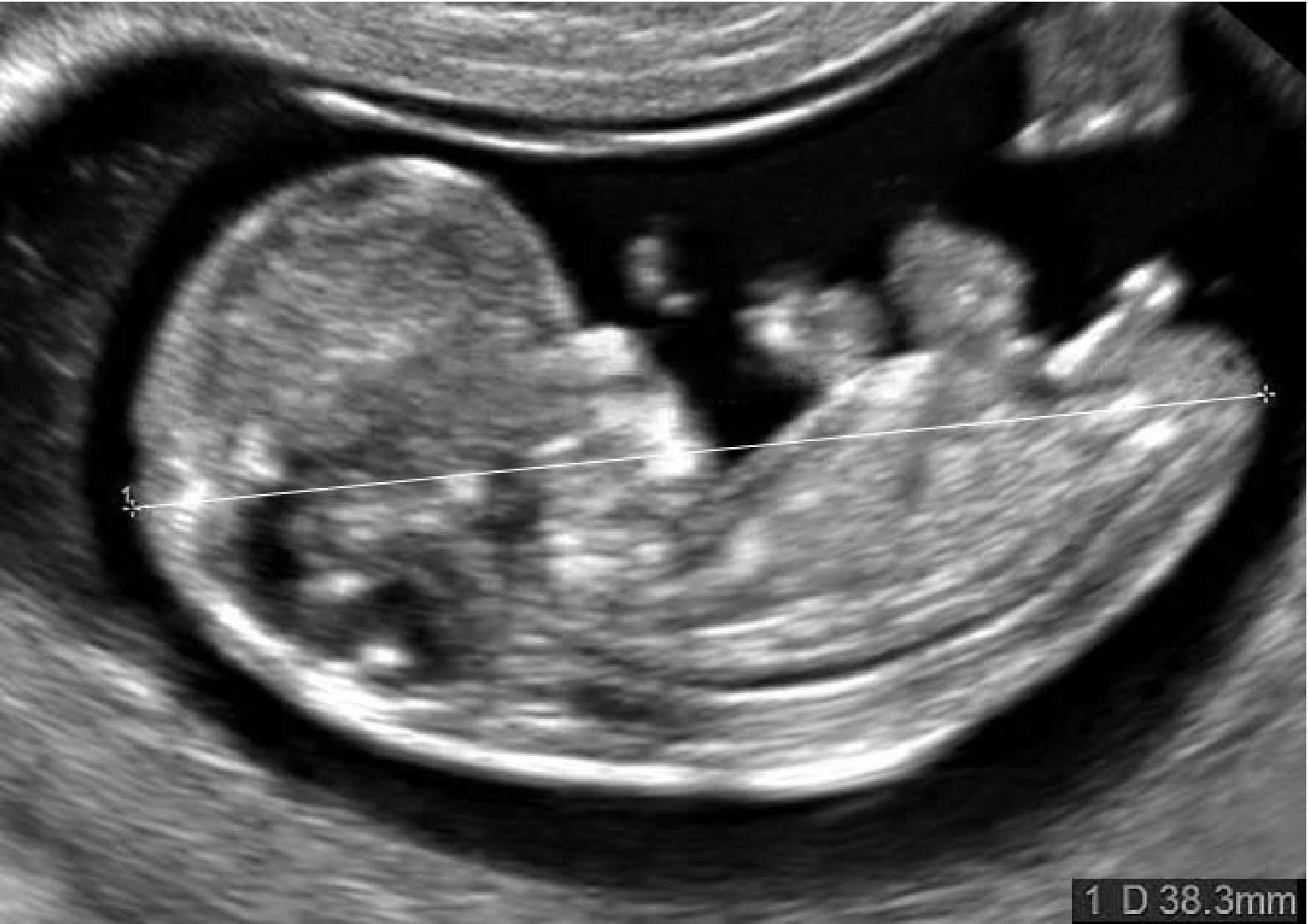}
&  \hspace{-0.5cm}     \includegraphics[width=6cm]{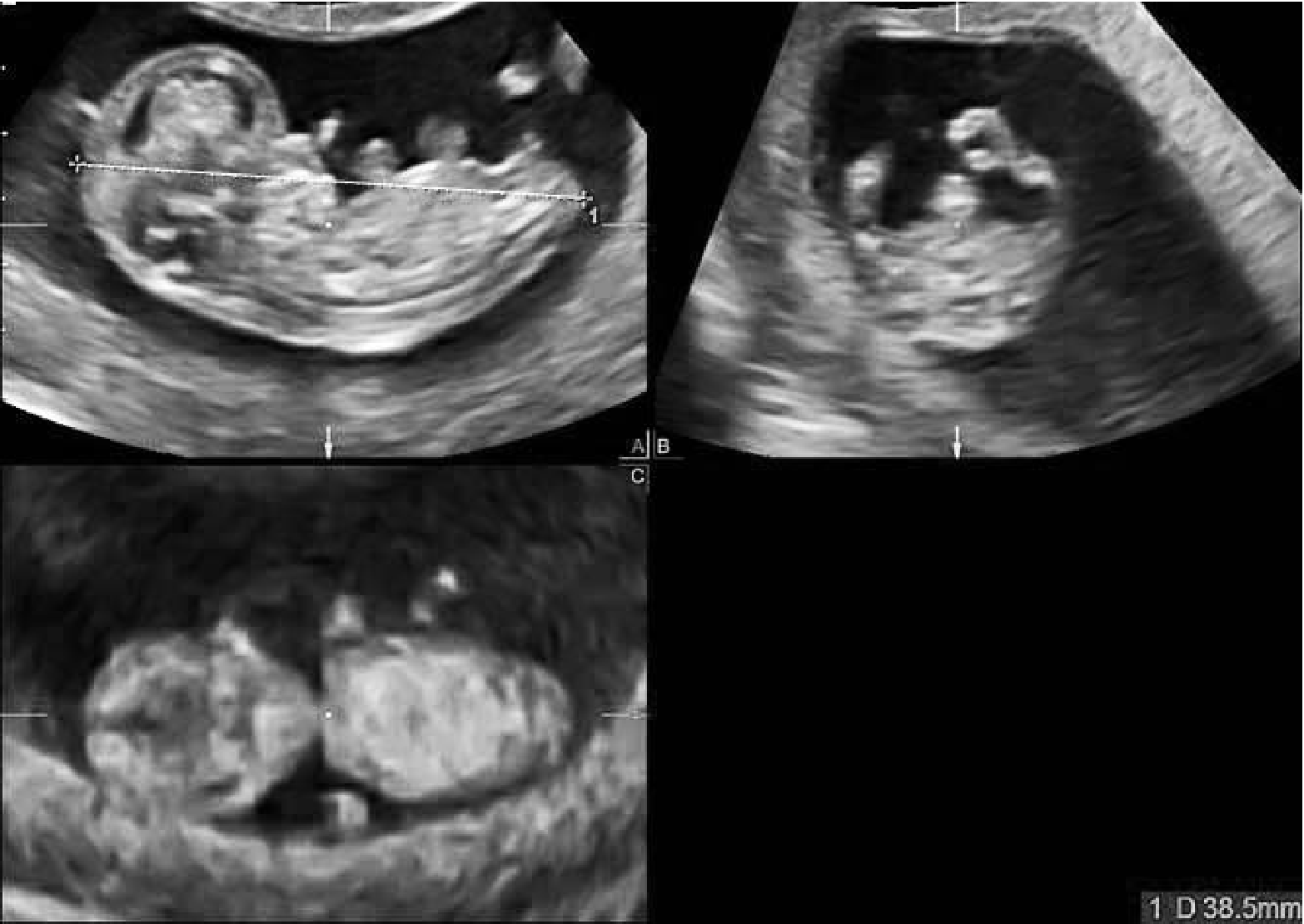}
\end{tabular}
 \vspace{-0.4cm}
     \caption{{\small \textit{Ultrasound pictures illustrating the embryonic length measurement. (a) Sagittal section of an 8 mm embryo. (b) Sagittal section of a 38 mm embryo. (c) Same embryo, sagittal section build with triplan mode from a 3D acquisition.}}}\label{fig1}
\end{figure*}

As several examinations were sometimes performed during the follow-up of pregnancies, only the measurement from the first examination was selected for inclusion in the datasets. Both in-vitro fertilization (B.B. sample) and spontaneous (M.C. sample) pregnancy studies were performed with devices from the same manufacturer: Kretz/General Electric. IVF sample (B.B. sample) was performed with different versions of Kretz Voluson 530, 730, Expert, General Electric E8 (from 1989-02-16 to 2010-09-17). Spontaneous pregnancy studies (M.C. sample) were performed with Kretz Voluson Expert V5 (from 2006-01-02 to 2007-01-01) and General Electric E8 (from 2008-01-02 to 2010-12-23)). The measures were performed and collected with the software MEDISPHINX\footnote{http://www.isisphinx.com/} which allows a precise collection of the clinical circumstances of the ultrasound examination. \\

The way embryonic lengths were measured was optimized for every examination, according to the size of the embryo and the prevailing technical conditions. Below 20 mm, the greatest length is a neck-rump measurement \cite{pexsters}. As often as possible, the measure tried to get closer to a sagittal section. This criterion is less important for small embryos. For many measures, a sagittal section was often obtained in 2D mode by vaginal or abdominal access. In case of difficulties, this sagittal section was be built with triplan mode from a 3D acquisition \cite{verburg} (Figure \ref{fig1}). Below 45 mm, the measure was almost essentially the one of an embryo in flexion. Beyond, we looked for a neutral position of the embryo with presence of an amniotic liquid triangle between the chin and the breastbone.

\begin{figure*}[!ht]
     \centering
\begin{tabular}{cc}
(a) & (b)\\
      \includegraphics[height=5cm]{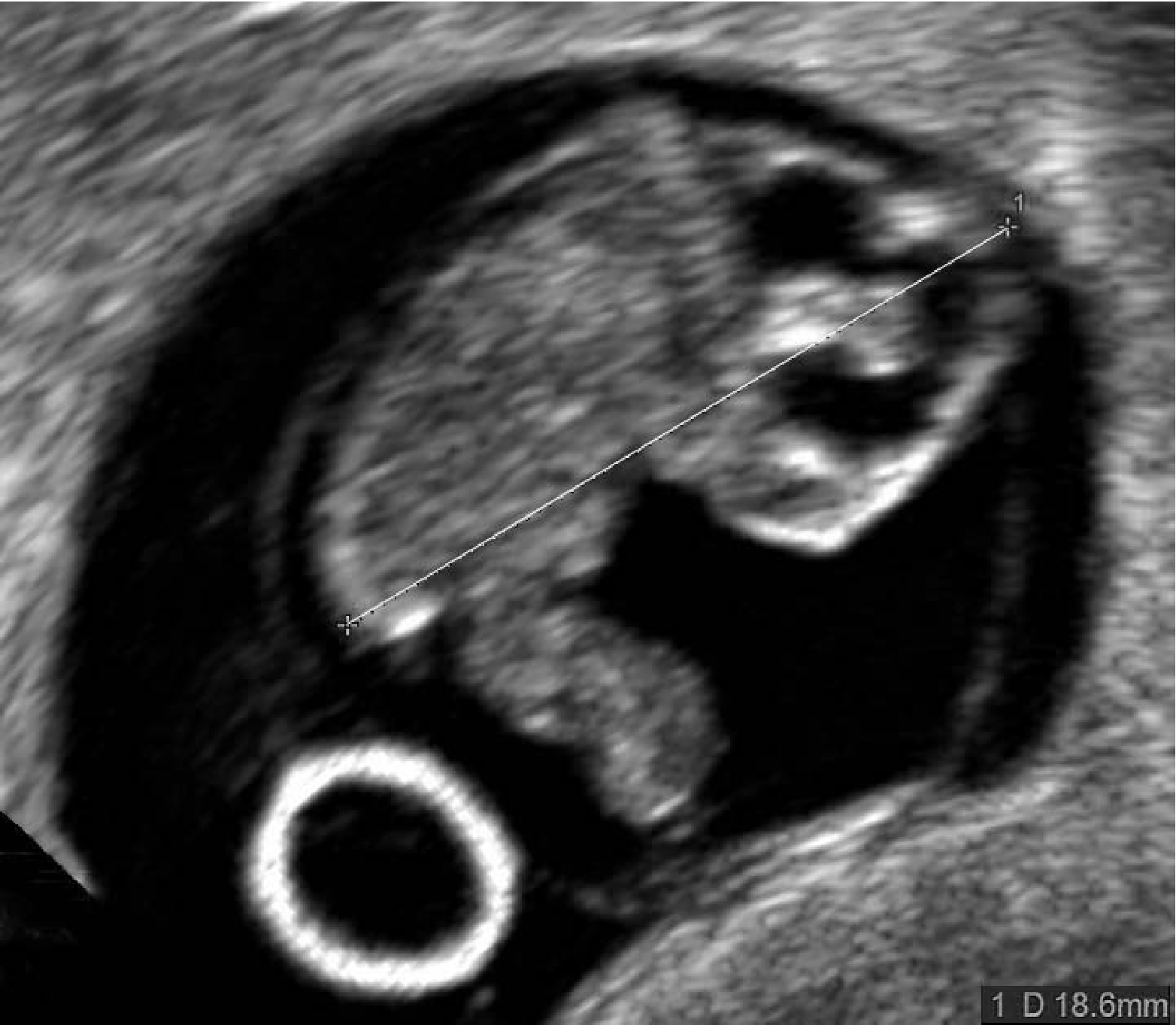} &       \includegraphics[height=5cm]{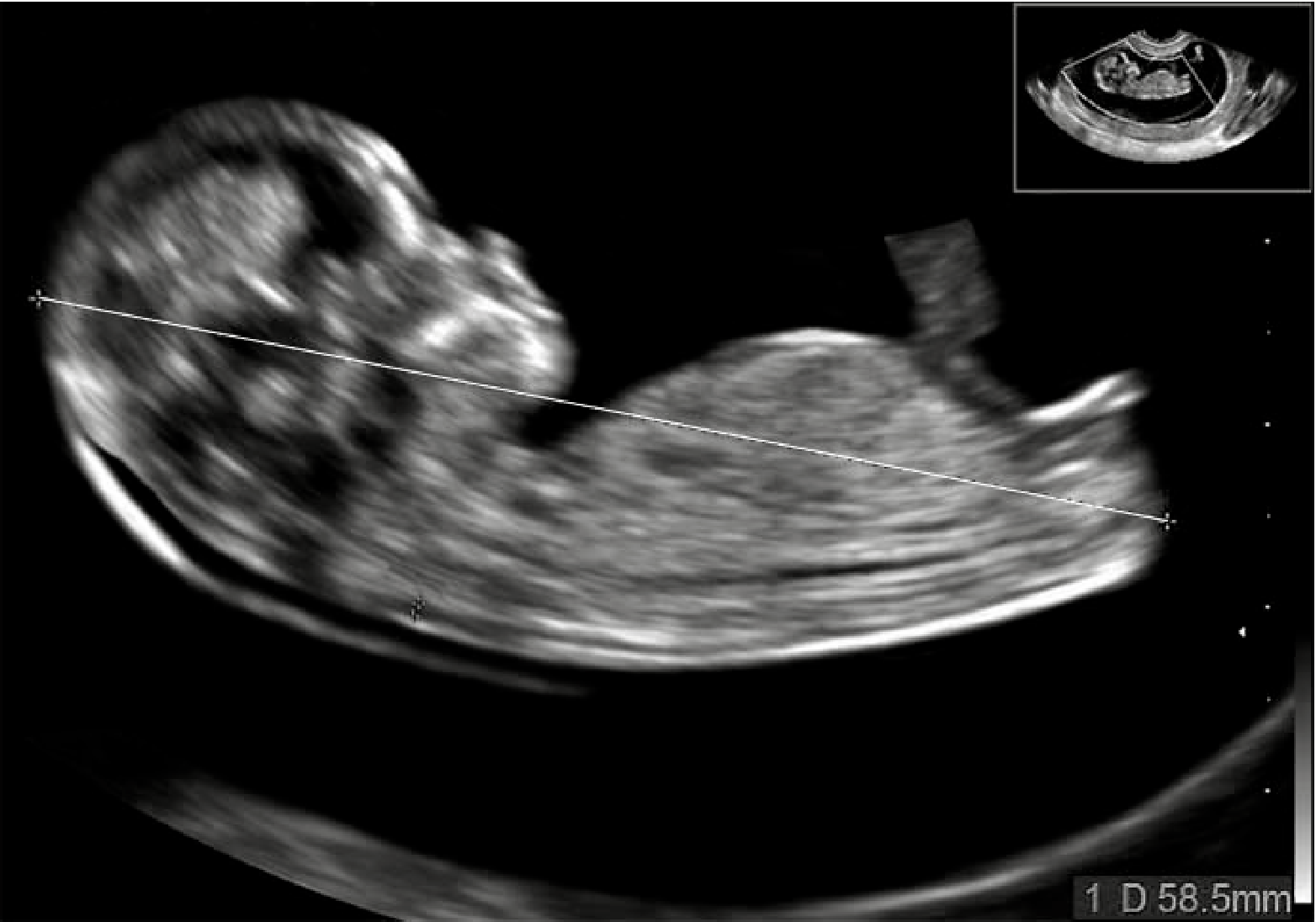}
\end{tabular}
     \caption{{\small \textit{(a) Ultrasound picture illustrating quality criteria for CRL below 45 mm (right, 19 mm). (b) Quality criteria beyond 45 mm (left, 59 mm).}}}\label{fig2}
\end{figure*}

The quality criteria researched for the measures of the CRL were as follows:\\
\noindent \underline{For CRL$<45$ mm (Figure \ref{fig2} (a))}
\begin{enumerate}
\item Sagittal section: visibility as soon as possible of the fourth ventricle and of the origin of the umbilical cord. Then, we added the rachis and the genital tuber.
\item Favorable locations of calipers: crown and rump were well defined, without edge contact, and the calipers were placed at the outer side of the interface (greatest length). Note that at this age, the biggest length is obtained by placing the cephalic caliper in the crown-neck junction \cite{pexsters}.
\item Zoom of the image: do that the image of the foetus filled at least 3/4 of the screen.
\end{enumerate}
\noindent \underline{For CRL$>45$ mm (Figure \ref{fig2} (b))}
\begin{enumerate}
\item Foetus in neutral position: presence of a triangle of amniotic liquid between the chin and the breastbone.
\item Sagittal section: ideally, visibility of hard plate or nasal bone, rachis and genital tuber (Figure 2).
\item Favorable locations of calipers: as in the previous case.
\item Zoom of the image: as in the previous case.
\end{enumerate}

To clean the data, we used non parametric tolerance limits \cite{murphy}\cite{somerville}. In this context if the consensus reference values for a Gaussian distributed parameter are usually those comprised between -2SD and +2SD (95\% of the whole population), for a non Gaussian distributed parameter as is probably the case here, there is no consensus range \cite{henny}. The 3rd and 97th or the 5th and 95th or the 10th and 90th percentiles are chosen as reference values according to the biological or physiological parameter under investigation, thus excluding either 6, 10 or 20\% of the population with extreme values \cite{must}\cite{zachmann}\cite{bjerre}. We performed a preliminary robust regression, that is not sensitive to outliers, and provided weights which were used to clean the data. This led us to exclude 8.4\% of the pregnancies with extreme values, 3.9\% above and 4.5\% below.

\subsection{Mathematical methods}

\noindent \textbf{Regressions:} Several regression models were tested. Numerical tests, based on comparisons of $R^2$ statistics, led us to describe CRL by polynomials of degree 2 in FA. Fisher's significance tests for the highest degree coefficients were performed and showed significant coefficients. The choice of degree 2 polynomials avoided overfitting of the model to the data and was in agreement with the settings of Robinson \cite{robinson73}, Verwoerd-Dikkeboom et al. \cite{verwoerd} or Pexters et al. \cite{pexsters}. Papaioannou et al. \cite{papaioannou} used a polynomial of degree 2 in FA to describe the square root of the CRL.\\
Whereas classical least square estimators assume a constant variance of the residuals, the discrepancy between predictions and observations in our case, clearly depended on the FA at the time of measure. Thus, we corrected the heteroskedasticity to account for the varying variance of the residuals and to obtain an equation for the latter. For this, we applied generalized least squares regressions\cite{greene}\cite{sheskin}. The modeling of the standard deviation was very important since our purpose was to further define Z-scores and normal parameters.\\
Our dataset exhibited outliers due among other to imprecision on the fecundation time, which could compromise the quality of the regression results. The robust regression methods we used are designed to be unsensitive to the presence of outliers \cite{maronna}\cite{rousseeuwleroy}. We used the \verb"lmRob" function from the \verb"R Cran" robust package \cite{wangrobust}. The study was completed by a modeling of the residuals' variance in time. Also, the observations corresponding to CRLs smaller or higher than 45mm were reweighted so that these two periods had equivalent importance in the regression.\\

\noindent \textbf{Validation of the results:} We considered two criteria to compare the results provided by the different regressions. The cross-validation reveals methods with good predictive power, whereas the study of the differences between predictions and measurements provides information on the accuracy of prediction.\\
The idea behind cross-validation is to separate the observations into two samples: the first one, the training sample, was used to obtain the regression formula that express the explained variable as a function of explanatory variables ; the explanatory variables of the second sample, the test sample, were used in these regressions to provide predictions of the explained variable. These predictions were compared with the true measurements. We chose a leave-one-out cross-validation that uses test samples of size 1 and repeats the procedure by successively excluding each observation from the training data. We did this for each regression that had been proposed (heteroskedastic, robust, Robinson's curve) and ranked the methods according to the number of times their predictions were the best ones.\\
To ensure that the methods did not perform too badly when they did not provide the best predictions, we also studied the distribution of the deviations between the predictions and the measurements.

\section{Results}

\subsection{Pregnancies from IVFs}

\begin{figure*}[!ht]
     \centering
\begin{tabular}{cc}
(a) & (b) \\
 \hspace{-0.5cm}    \includegraphics[width=6cm]{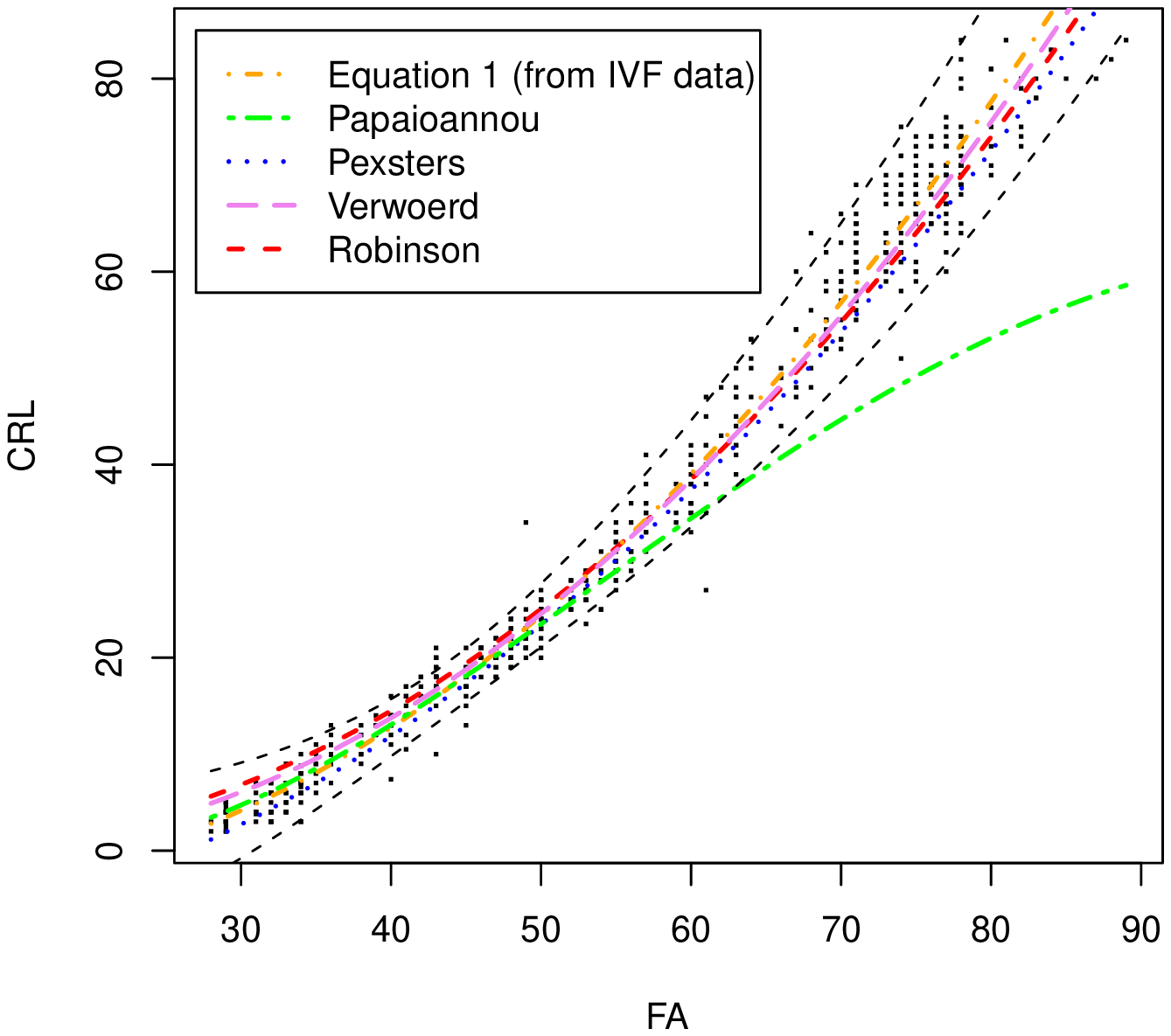}
&  \hspace{-0.5cm}   \includegraphics[width=6cm]{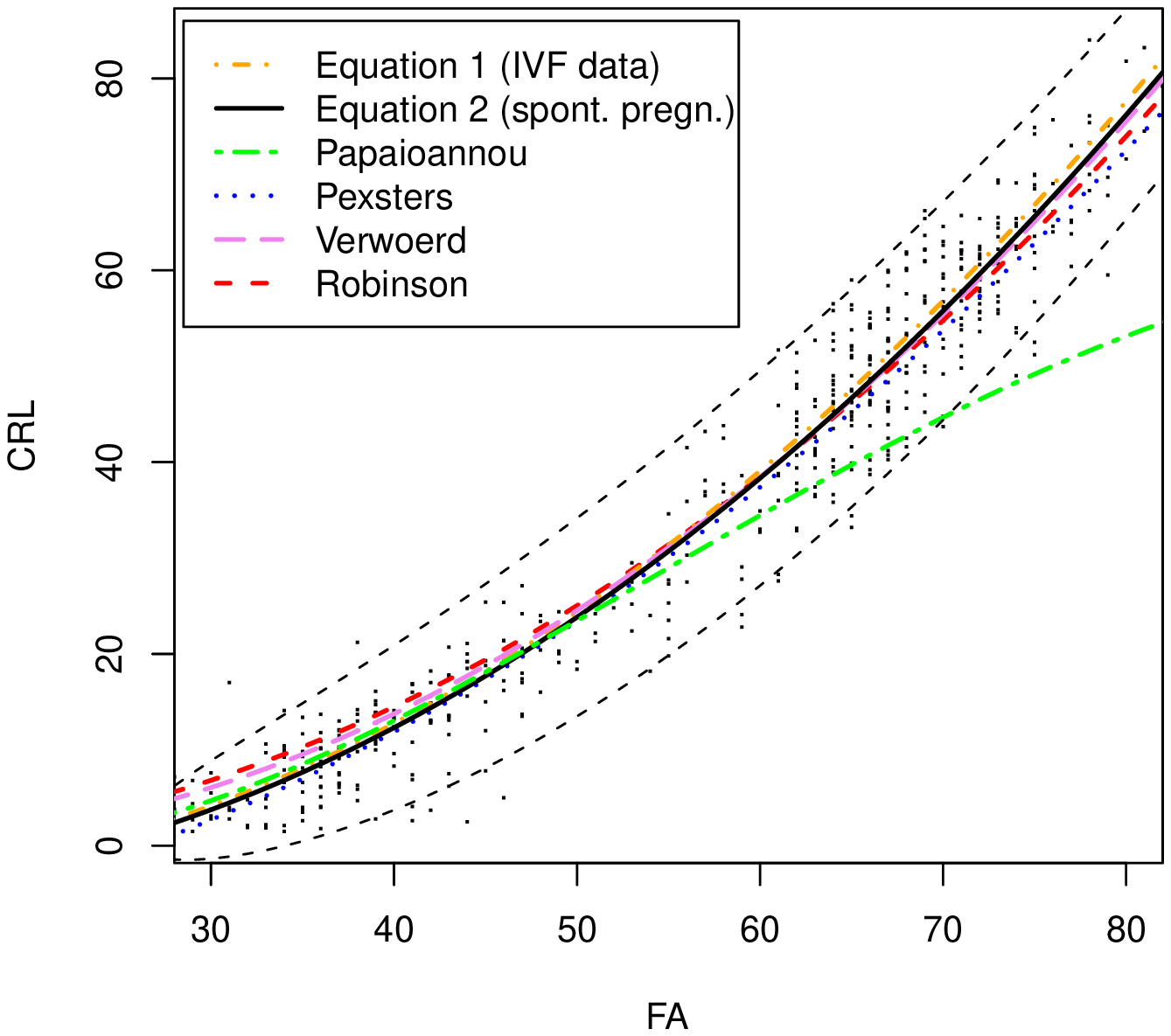}
\end{tabular}
 \vspace{-0.4cm}
     \caption{{\small \textit{(a) Predictions of the CRL values with equation \eqref{eqFIVrob1:LCC=f(JV)} and with Papaioannou's, Pexsters's, Verwoerd-Dikkeboom's and Robinson's models. IVF data are represented. The confidence interval associated with equation \eqref{eqFIVrob1:LCC=f(JV)} and obtained with our IVF data is represented in black dotted line.\newline (b) Predictions of the CRL values using equation \eqref{eqFIVrob1:LCC=f(JV)} (from our IVF data), equation \eqref{eq:LCC-O2=f(JV)} (from our spontaneous pregnancies data) and using Papaioannou's, Pexsters', Verwoerd-Dikkeboom's, Robinson's models. The data from spontaneous pregnancies are represented. The confidence intervals associated with \eqref{eq:LCC-O2=f(JV)}, with $\kappa=1.96$, are drawn in dotted black lines. }}}\label{fig:predictions1}
\end{figure*}

Out of the 402 pregnancies, there were 78 (19.40\%) twin pregnancies and 8 (1.99\%) triple pregnancies. Preliminary tests (Wilcoxon and Chow, data not shown) showed that multiple pregnancies did not required to be separated from single pregnancies and that the measures made in Nice and Monaco could be considered simultaneously.

Heteroskedastic and robust regressions were realized. We performed cross-validation with the obtained regressions (see Table \ref{table:VC1}). Globally, the robust regression gave the best prediction in 64.56\% cases:
\begin{align}
& \CRL= - 3.3108 - 0.2087\ \FA + 1.5250\,10^{-2}\ \FA^2,\label{eqFIVrob1:LCC=f(JV)}\\
 & \mbox{with }R^2=73.79\% \nonumber\\
& \sigma^2=46.2354 - 2.0194 \ \FA+0.0230\ \FA^2,\nonumber
\end{align}Tabulated values are given in the electronic supplementary materials (ESM).\\

We used robust tests that generalize the Wald test for linear hypothesis\cite{maronna}. The null assumptions that the coefficients in \eqref{eqFIVrob1:LCC=f(JV)} were equal to the coefficients of Robinson, Pexsters or Verwoerd-Dikkeboom models were all rejected with p-values smaller than $0.01$.\\

We saw from the cross-validation (Table \ref{table:VC1}) that the CRLs predicted with our model were closer to the true values in 64.56\% of the cases when compared with Robinson's prediction (67.72\% of the cases when compared with Papaioannou's curve, 63.68\% when compared with Pexsters' curve, 54.56\% when compared with Verwoerd-Dikkeboom's curve).
Predictions errors, with \eqref{eqFIVrob1:LCC=f(JV)}, had also a smaller standard deviation (3.43 mm).

\begin{table*}[ht]
\begin{center}
\begin{tabular}{|c|ccccc|}
\hline
& Equation \eqref{eqFIVrob1:LCC=f(JV)}  & Robinson's  & Papaioannou's & Pexsters's & Verwoerd-Dikkeboom's\\
& (from IVF data) &  curve \eqref{equationRobinson} &  curve \ref{eq:papa} & curve  \eqref{equationpexsters}& curve \eqref{equationverwoerd}\\
\hline
 \multicolumn{6}{|c|}{Number of best performances}\\
\hline
Frequency & 192 (33.68) & 85 (14.91) & 63 (11.05) & 156 (27.37) & 74 (12.98)\\
(Percentage) (\%)  & 368 (64.56) & 202 (35.44)& * &  *  & *\\
 & 386 (67.72) & * & 184 (32.28) &  *  & *\\
 & 363 (63.68) & * & * & 207 (36.32)  & *\\
 & 311 (54.56) & * & * & * & 259 (45.44)\\
 \hline
 \multicolumn{6}{|c|}{Absolute error}\\
 \hline
Min  & 0.01702 & 0.0350 & 0.0193 &  0.0020  & 0.0086\\
Median  & 1.7690 & 2.1670 & 2.9560 & 2.1380 & 1.9150\\
Mean  & 2.4650 & 2.9400 & 7.9240 & 3.0470 & 2.6530\\
Max  & 14.9200 & 14.1100 & 32.3800 & 15.4400 & 12.9900\\
\hline
 \multicolumn{6}{|c|}{Error}\\
\hline
Min absolute error & -10.9900 & -14.1100 & -32.3800 & -15.4400 & -12.75\\
Median absolute error & -0.0879 & 0.4710 & -2.5120 & -1.8900 & 0.2488\\
Mean absolute error & 0.0481 & -0.3087 & -7.2090 & -2.170 & -0.2373\\
Max absolute error &  14.9200 & 12.9700 & 8.5150 & 11.9100 & 12.990\\
Std dev. & 3.4393 & 3.8680 & 8.9772 & 3.5432 & 3.587\\
\hline
\end{tabular}
\caption{\textit{In the first part of the table, we compared the different equations by cross-validation. In line 1, we computed the predictions for all the models and count, for each method, the number of times it gave the best prediction. In lines 2 to 5, we compared equation \eqref{eqFIVrob1:LCC=f(JV)} with each of the other model (for instance, in line 2, we compared the predictions of equation \eqref{eqFIVrob1:LCC=f(JV)} and of Robinson's model and indicated how many times each of them gave the best prediction). In the second and third parts of the table, summary of the absolute error and of the signed error were computed.
The data used for the cross-validation were the IVF measures. Our equation \eqref{eqFIVrob1:LCC=f(JV)} provided the best predictions. Papaioannou's curve was efficient for GA under 75 days (FA under 61 days) but not for larger embryos (see ESM).}}\label{table:VC1}
\end{center}
\end{table*}

\subsection{Pregnancies after spontaneous conception}

Using the same methodology as reported above, we performed heteroskedastic and robust regressions with the CRL measures from spontaneous pregnancies.
\begin{align}
& \CRL= - 4.1212 - 0.1824\ \FA + 0.0148\ \FA^2,  \label{eq:LCC-O2=f(JV)} \\
& \mbox{with }R^2=78.48\%\nonumber\\
& \sigma^2=-53.1054+2.5634\ \FA-0.0189\ \FA^2,\nonumber
\end{align}The CRL 95\% confidence interval was also determined:
\begin{multline*} CI=[ - 4.1212 - 0.1824\ \FA + 0.0148\ \FA^2 \\
 \pm \kappa \times \big(-53.1054+2.5634\ \FA-0.0189\ \FA^2\big)^{1/2}].
\end{multline*}The constant $\kappa$ was calibrated so that 95\% of the observations were included in the confidence interval, thus $\kappa=1.85$.

This regression \eqref{eq:LCC-O2=f(JV)} was compared to Robinson's, Papaioannou's, Pexsters', Verwoerd-Dikkeboom's equations and to equation \eqref{eqFIVrob1:LCC=f(JV)} (obtained with IVFs). Robust Wald-type tests \cite{maronna} showed that with the variance of our observations, our equation \eqref{eq:LCC-O2=f(JV)} could be considered as significatively different from Robinson's regression (p-value=0.0261). The equations of Pexsters, Verwoerd-Dikkeboom and equation \eqref{eqFIVrob1:LCC=f(JV)} were not statistically different from equation \ref{eq:LCC-O2=f(JV)} (p-values of 0.0768, 0.3969 and 0.8189 respectively), although the cross-validation and the study of the errors between predictions and observations (see Table \ref{tab:equationMC-comparaison}) showed that there were closest predictions with our equation \ref{eq:LCC-O2=f(JV)}.

\begin{table*}[!ht]
\begin{center}
\begin{tabular}{|c|cccccc|}
\hline
&  Equation \eqref{eq:LCC-O2=f(JV)} &  Equation \eqref{eqFIVrob1:LCC=f(JV)} & Robinson & Papaioannou &  Pexsters & Verwoerd \\
& (from spont. pregn.)  & (from IVF)  &
\eqref{equationRobinson} & \eqref{eq:papa}& \eqref{equationpexsters} & \eqref{equationverwoerd}\\
\hline
 \multicolumn{7}{|c|}{Number of best performances}\\
\hline
Frequency & 269 (52.44) & 244 (47.56) & * & * & * & *\\
(Percentage) (\%)  & 286 (55.75) & * & 227 (44.25) & * & * & *\\
& 338 (65.89) & * & * & 175 (34.11) & * & * \\
& 283 (55.17) & * & * & * & 230 (44.83) & *\\
& 258 (50.29) & * & * & * & * & 255 (49.71)\\
& * & 282 (54.97) & 231 (45.03) & * & * & *\\
 & * & 337 (65.69) & * &  176 (34.31) & * & *\\
 & * & 266 (51.85) & * & *& 247 (48.15) & * \\
  & * & 266 (51.85) & * & *& *& 247 (48.15) \\
\hline
 \multicolumn{7}{|c|}{Absolute error}\\
 \hline
Min  & 0.0055 & 0.0137 & 0.0358 & 0.0450 & 0.0100 & 0.0232\\
Median & 3.3200 & 3.1800 & 3.5510 & 5.4840 & 3.3620 & 3.3070\\
Mean  & 4.1150 & 4.1430 & 4.2950 & 7.6410 & 4.2330 & 4.1610\\
Max  & 14.6000 & 15.8800 & 15.8600 & 32.3800 & 15.4400 & 15.2000\\
\hline
 \multicolumn{7}{|c|}{Error}\\
\hline
Min  & -12.4800 & -12.1200 & -14.1100 & -32.3800 & -15.4400 & -12.7500\\
Median & -0.07181 & 0.6235 & 0.8910 & -3.7620 & -1.2900& 0.7229\\
Mean & 0.1997 & 1.0320 & 0.6668 & -5.6230 & -1.0830 & 0.7166\\
Max  & 14.6000 & 15.8800 & 15.8600 & 14.5600 & 13.7000 & 15.2000\\
Std dev. & 5.2175 & 5.2102 & 5.4044 & 8.3903 & 5.2563 & 5.2565\\
\hline
 \multicolumn{7}{|c|}{Absolute relative error (Percentages)}\\
\hline
Median & 11.0700 & 11.1200 & 10.2200 & 19.6100 & 11.4400 & 10.5000\\
Mean  & 24.9900 & 26.0400 & 32.0300 & 31.5900 & 24.2500  & 29.1900\\
\hline
\end{tabular}
\caption{\textit{Comparison of the predictions of Equations \eqref{eq:LCC-O2=f(JV)} and \eqref{eqFIVrob1:LCC=f(JV)} with Robinson's, Pexters's and Verwoerd-Dikkeboom's equations, using the reweighted sample of 513 women with regular cycles. In the first part of the table, cross-validation with the data from spontaneous pregnancies was conducted, as in Table \ref{table:VC1} ; in the second and third parts, summaries of the absolute and signed errors are provided.}}\label{tab:equationMC-comparaison}
\end{center}
\end{table*}

Using spontaneous pregnancies data, for FA = 28 days, equation \eqref{eq:LCC-O2=f(JV)} (from spontaneous pregnancies) predicted CRL = 2.39 mm, equation \eqref{eqFIVrob1:LCC=f(JV)} CRL=1.98 mm, thus a -0.41 mm difference with IVF equation \eqref{eqFIVrob1:LCC=f(JV)}. Likewise for FA=28 days Robinson's curve predicted CRL = 3.23 mm (+0.84 mm) with equivalent to an overestimation of 1 day by the latter, Papaioannou's curve predicted CRL = 1.86 mm (-0.53 mm), Pexsters' curve predicted CRL=1.16 mm (-1.23 mm) and Verwoerd-Dikkeboom's curve predicted CRL=2.51 mm (-0.12 mm).\\
For FA=70 days, equation \eqref{eq:LCC-O2=f(JV)} (from spontaneous pregnancies) predicted CRL=55.77 mm. We found a difference of 1.03 mm with the equation \eqref{eqFIVrob1:LCC=f(JV)} (IVF equation), -1.00 mm with Robinson's curve, -11.11 mm with Papaioannou's curve, -2.03 mm with Pexsters' curve and -0.35 mm with Verwoerd-Dikkeboom's curve.\\

For CRL $<$ 20 mm, no significant difference was noticed between equation \eqref{eqFIVrob1:LCC=f(JV)} (built on IVF) and equation \eqref{eq:LCC-O2=f(JV)} (built on spontaneous pregnancies) , p = 0.8713. For CRL = 20 mm the mean difference between the predictions of equation 1 and equation 2 was 0.43 mm (median of 0.42 mm and standard deviation of 0.039 mm, p-value$<$0.05) and fits with less than 1 day difference for foetal age. Thus, on this period, datations predcited by these equations were either similar or negligible (see datation tables in ESM). However, for CRL = 20 mm, the deviation between equation \eqref{eqFIVrob1:LCC=f(JV)} and equation \eqref{eq:LCC-O2=f(JV)} slightly increased. Graphically (Fig. \ref{fig:predictions1} (b)), the equation \eqref{eqFIVrob1:LCC=f(JV)} (built on IVF) over-estimated the CRL.\\

A t-test showed that, for the spontaneous pregnancies, the proportion of over-estimations by the equation \eqref{eqFIVrob1:LCC=f(JV)} (built on IVF) was statistically higher than when equation \eqref{eq:LCC-O2=f(JV)} (built on spontaneous pregnancies) was used (p-value= 0.0330). The mean difference between the CRL prediction by equation \eqref{eqFIVrob1:LCC=f(JV)} and equation \eqref{eq:LCC-O2=f(JV)} was 0.94 mm, with a median of 0.94 mm and a standard deviation of 0.223 mm.

\subsection{Predicting FA with the ultrasound measure of CRL}

Up to now, the determination of FA from the ultrasound measure of CRL is based on Robinson's equation \cite{robinson75} or on the equation by Papaioannou et al. \cite{papaioannou} (see equations in the ESM). As these authors, we opted to build a mathematical model from our data predicting FA from CRL. Indeed, in a stochastic setting, inverting the equations linking measures of CRL to FA may not be optimal. We built mathematical models from our data using the same regression methods and validation procedures as described in the previous sections.\\

\begin{figure}[!ht]
     \centering
\includegraphics[width=6cm]{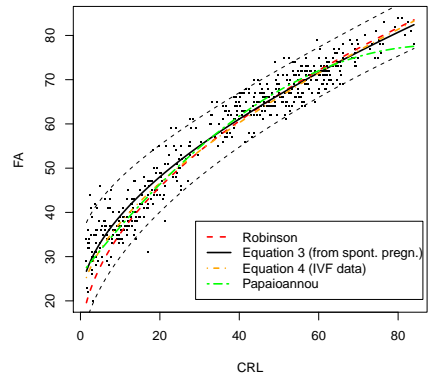}
 \vspace{-0.4cm}
     \caption{{\small \textit{Predictions of the foetal age (time since fecundation) from the measurements of the CRL using the equations found with various methods. We represented data from the sample of spontaneous pregnancies.
     }}}\label{fig:predictions}
\end{figure}

This time, the heteroskedastic regression was shown to be more accurate when compared with the equation obtained by robust regression (with a cross-validation and a study of the prediction error, data not shown). These equations were:
\begin{align}
& \FA=19.1732 + 6.0266\ \sqrt{\CRL} + 0.0955\ \CRL, \label{eq:JV-HT}\\
& \mbox{with }R^2=99.71\% \nonumber\\
& \sigma^2=33.1275-4.9440 \ \sqrt{\CRL}+0.2270\ \CRL.\nonumber
\end{align}and showed relatively small heteroskedasticity.\\

Using the IVF data, we obtained the following equation for predicting FA from CRL:
\begin{align}
& \FA = 18.0739+ 5.6925 \sqrt{\CRL}+  0.1549 \CRL, \label{eq:JV-FIV}\\
& \mbox{with }R^2=99.87\% \nonumber\\
& \sigma^2= 7.3281- 1.6397 \sqrt{\CRL}+0.1688 \CRL.\nonumber
\end{align}

Fisher's tests with heteroskedastic corrections showed that the coefficients of \eqref{eq:JV-HT} and those of Robinson \cite{robinson75} were statistically different (p-value=0.0321). The difference between the coefficients of \eqref{eq:JV-HT} and \eqref{eq:JV-FIV} was not statistically significant (p-value=0.9943). Papaioannou et al. \cite{papaioannou} used a polynomial of degree 2 to express FA as a function of CRL and CRL$^2$ thus the coefficients could not be compared to the ones of \eqref{eq:JV-HT}.\\

We compared the equation \eqref{eq:JV-HT} (built on the spontaneous pregnancies), with the equation \eqref{eq:JV-FIV} (built on IVFs), Robinson's equation and Papaioannou's equation (see Table \ref{tab:equationMC-comparaisonJV}). This comparison showed that Equation \eqref{eq:JV-HT} and Papaioannou's equations had similar predictive powers and equivalent precision in the predictions. Both outperformed Robinson's equation, but the latter still provided correct predictions for CRL larger than 45 mm, with an average error of 3.5 days (against 3 days for Equation \eqref{eq:JV-HT} and Papaioannou's equation, see Figures \ref{fig:predictions}).

\begin{table*}[!ht]
\begin{center}
\begin{tabular}{|c|cccc|}
\hline
Regression & Equation \eqref{eq:JV-HT} &    Equation \eqref{eq:JV-FIV} & Robinson & Papaioannou et al. \\
& (Spont. pregn.) & (IVF) & \eqref{equationJVRobinson} & \eqref{eq:papajv}\\
\hline
 \multicolumn{5}{|c|}{Number of best performances}\\
\hline
Frequency  & 262 (51.07) & 251 (48.93) &  * & *\\
(Percentage) (\%) &  &  &  & \\
&  304 (59.26) & * &   209 (40.74)& *\\
 & 251 (48.93) & * & * & 262 (51.07)  \\
  & * & 281 (54.78) & 232 (45.22) & * \\
  & * & 246 (47.95) & * & 267 (52.05)\\
  & * & * & 208 (40.55) & 305 (59.45)\\
\hline
 \multicolumn{5}{|c|}{Absolute error}\\
 \hline
Min absolute error & 0.0071 & 0.0033 & 0.0100 & 0.0075\\
Median absolute error & 2.5410 & 2.3220 & 2.4310 & 2.2660\\
Mean absolute error & 2.9180 & 2.9310 & 3.3550  & 2.8490\\
Max absolute error & 15.1400 & 16.5400 & 21.5400 & 15.3400\\
\hline
 \multicolumn{5}{|c|}{Error}\\
 \hline
Min error & -15.1400 & -16.5400 & -21.5400 & -15.34\\
Median error & 0.2091 & -0.4440 & -0.8924 & -0.2494\\
Mean error & -0.0023 & -0.8233 & -1.4820 & -0.5261\\
Max error & 14.7000 & 13.1800 & 11.9300 & 12.60\\
Std dev. & 3.7200 & 3.7352 & 4.3729 & 3.6717\\
\hline
 \multicolumn{5}{|c|}{Relative error (Percentages)}\\
 \hline
Median absolute error & 4.1790 & 4.2970 & 4.3940 & 4.1870\\
Mean absolute error & 5.9990 & 5.8790 & 6.9990 & 5.6700\\
\hline
\end{tabular}
\caption{\textit{Comparison of the predictions of the different equations for $\FA$ based on observations of the $\CRL$. The first part of the table describes the cross-validation with the data from spontaneous frequencies. The first line gives for each model the number and percentage of times they provided the best prediction, when compared all together. The lines 2 to 5 compare the models two by two. The second, third and fourth part of the table provide summary statistics for the absolute error, the signed error and the relative error. }}\label{tab:equationMC-comparaisonJV}
\end{center}
\end{table*}

\subsection{Optimal period to predict the FA with CRL}

From equation \eqref{eq:JV-FIV}, the foetal age at which the width of the confidence interval of FA predicted from the CRL was the smallest, was obtained for CRL=23.5950 mm, corresponding to FA=49.38 days. The CRL standard deviation at this time was estimated to be 1.8292 mm. This corresponds to the period where there is a trade-off between measuring large dimensions and adequate position of the embryo. Around FA=50 days, the embryo is indeed quite curved in shape and motionless.

\subsection{Breakpoint}

\begin{figure}[!ht]
     \centering
     \includegraphics[width=6cm]{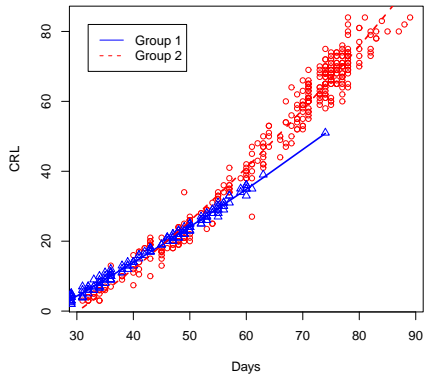}
     \caption{{\small \textit{Estimation of mixed linear regression with the Flexmix package, with the two clusters that have been identified. IVF data are used.}}}\label{fig:regFIV-flexmix}
\end{figure}

As the graphical representation of CRL by FA (Fig. \ref{fig:predictions1} (a)) evoked a breakpoint in foetal growth, instead of using polynomials of degree 2 to model the CRL as a function of FA, we searched to reproduce the curvature of the graph by modeling a breakpoint between two subperiods. We looked for breakpoints from mixed regressions and used the package \verb"flexmix" \cite{grunleisch}. We obtained two groups (Fig. \ref{fig:regFIV-flexmix})that may be modeled respectively with:
\begin{align*}
& \CRL= - 21.15 + 0.7642\ \FA +2.820\ \FA^2\nonumber\\
& \CRL= - 28.1408+ 0.7106\ \FA + 7.364\ 10^{-3} \FA^2,
\end{align*}with $R^2$-statistics of 99.15\% and 97.36\%. Both equations disclosed a breakpoint at FA=45.56 days.

The methods distinguished two samples of respectively 211 and 359 observations, which differed from their FA, the first sample being composed of small FAs.

\section{Discussion}

Robinson's curve remains a historical benchmark that is still widely used. Since 1973, many other studies have been performed. Our first purpose was to propose new equations for the modeling of CRL as a function of FA using modern equipment and following the standardized procedures for ultrasound measurements in the context of Down syndrome's detection in a general population.
We obtained two equations. For IVF data, equation \eqref{eqFIVrob1:LCC=f(JV)} and equation \eqref{eq:JV-FIV} respectively gave the predictions of CRL by FA and of FA by CRL. Likewise with spontaneous pregnancies, we obtained equation \eqref{eq:LCC-O2=f(JV)} and equation \eqref{eq:JV-HT}.\\
The accuracy of these new main equations we propose relies on several factors. First, each of our datasets (from IVFs and from spontaneous pregnancies) was obtained by a single operator using the same manufacturer devices: Kretz/General Electric. The computerized data collection was done on interfaces that had been specially developed for this work. Measurements were optimized by the choice of the best method, transabdominal or transvaginal scan, for every examination. The difficult measures benefitted from 3D triplan reconstructions. Moreover, we used an explicit methodology to clean our data. The statistical accuracy of our equation was established by a cross-validation and by studying the errors of the different predictions. Comparisons between the various predictions showed that for CRLs between 45 and 85 mm, the differences between ``ancient'' and ``recent'' charts were small. On the other hand, the recent equations benefitted from a larger number of data for the initial period, collected by vaginal access. The new equations we propose are more precise, and allow moreover to follow up in time embryos that are measured with CRL smaller than 45 mm. The beginning of the curve is much more accurate and is more in adequation with the practice, for instance when considering the predicted size after 28 days since conception.\\
Moreover, we provided models for the evolution of the residuals, which allows for building confidence intervals and Z-scores. As a byproduct we showed that the best period to predict FA is to measure embryo around day 50, when its size is around 24 mm.\\

For CRL $<$20mm there is not difference between equations built on IVF data and on spontaneous pregnancies data. This objectively suggests that the age at which an embryo of 1 to 2mm becomes visible in the gestational sac, as well as the embryonic growth for foetuses between 1 and 20 mm, are in average independent from the mode. For CRL$\geq$20mm the small difference between both models increased suggesting that IVF embryos could have a slightly higher growth rate than spontaneous pregnancies embryos as was already reported \cite{dias}\cite{MacGregor}.

Finally, models where the CRL is expressed as a polynomial of degree 2 are convenient for practical uses. However, searching a model with a break point showed that there is a change in the growth regime around day 46, when the CRL is around 20 mm. This could correspond to the beginning of the maternal connection.\\

\bigskip
\noindent \textbf{Acknowledgements} We are grateful to Christian Duroux for his contribution to the optimization of the obstetric part of MEDISPHINX. Prior to this work, M.C. and V.C.T. have supervised internship trainings of Anthony Cousien and Anthony Létendart. We thank Dr. Laurent Salomon and Florian Odor for interesting discussions. Dr. Omolade Alao read and corrected the manuscript.

{\footnotesize

}
\clearpage

\appendix
\onecolumn
\noindent {\Large\textbf{Electronic Supplementary Materials}}

\section{Equations of Robinson (1973, 1975)}

Robinson \cite{robinson73} calibrated CRLs from 334 measures of natural pregnancies with known and significant last menstruations:
\begin{equation}
\CRL=7.295-0.6444\ \GA+0.0144\ \GA^2,\label{equationRobinson}
\end{equation}where we recall that $\GA=\FA+14$ represents the gestational age. A corrective factor of (1 mm +3.7\%) was substracted by par Robinson to account for the calibration of his machine.\\

In \cite{robinson75}, Robinson established the following equation for computing the gestational age from measurements of the CRL.
\begin{equation}
\GA=8.052\sqrt{\CRL}+23.73.\label{equationJVRobinson}
\end{equation}

\section{Equation of Papaioannou et al. (2010)}

Papaioannou et al. \cite{papaioannou} obtain a curve for the $\CRL$ as a function of $\GA$ for $\GA$ between 40 and 75 days.
\begin{equation}
\sqrt{\CRL}=-6.662367  + 0.246741 \ \GA - 0.001046  \GA^2.\label{eq:papa}
\end{equation}
Their curve is correct for $\GA\leq 75$ days, but from the negative coefficient of $\GA^2$, we can guess that the fit for larger time will deteriorate (see Fig. \ref{fig:predictions}).\\

They could also predict the GA as a function of the CRL.
\begin{equation}
\GA = 39.811963 + 1.155896 \  \CRL - 0.006429 \ CRL^2.\label{eq:papajv}
\end{equation}

\section{Equation of Pexters et al. (2010)}

Using a large population of pregnancies, Pexsters et al. \cite{pexsters} proposed a new curve to re-estimate the CRL curve from 5.5 weeks' gestation and obtained:
 \begin{equation}
 \CRL=-9.09-0.26\ \GA+0.012\ \GA^2.\label{equationpexsters}
 \end{equation}

\section{Equation of Verwoerd-Dikkeboom et al. (2010)}

Verwoerd-Dikkeboom et al. \cite{verwoerd} use 3D measurements from 6 to 14 weeks GA in 32 pregnancies to calibrate an equation of the CRL.
\begin{equation}
\CRL=9.0963-0.751165\ \GA+0.015508\ \GA^2.\label{equationverwoerd}
\end{equation}
\begin{equation}
\sigma^2=0.2814-0.006087\ \GA+ 0.000043\ \GA^2.
\end{equation}

\section{Robust regression for the prediction of FA as a function of CRL}

The robust regression gave, on our IVF data:
 \begin{align}
& \FA = 17.8994+ 5.7617\ \sqrt{\CRL}+  0.1471 \ \CRL,\label{JV:lmRob}\\
& \sigma^2 = 7.3281-1.6397\  \sqrt{\CRL}+0.1688  \ \CRL.\nonumber
\end{align}Cross validation showed that on IVF data, the regression with heteroskedastic regression \eqref{eq:JV-FIV} gave the best prediction. Compared with Robinson's regression, it gave the best prediction in 343 cases out of 570 (60.18\%).\\

On spontaneous pregnancies data, we obtained:
\begin{align}
& \FA = 19.2702+5.7804 \sqrt{\CRL} +0.1271 \CRL , \label{eq:JV-robust}\\
& \sigma^2=41.8353-8.3486 \, \sqrt{\CRL}+0.5198 \CRL.
\end{align}
Cross-validation showed that on spontaneous pregnancies data, the heteroskedastic regression presented in \eqref{eq:JV-HT} gave better prediction in 263 (51.27\%) cases versus 250 (48.73\%) cases for the Robust regression. However, in term of precision, these two methods were comparable. For the heteroskedastic (resp. robust) regression, the mean absolute error was 2.9180 (resp. 2.9120), the median absolute error was 2.5410 (resp. 2.5000) and the standard deviation was 3.7199 (resp. 3.7169).

\section{Characteristics of the parents}

Characteristics of parents of spontaneous pregnancies data are presented the characteristics in Table \ref{tab:equationMC-comparaisonJV}.
\begin{table*}[!h]
\begin{center}
\begin{tabular}{|c|cc|cc|}
\hline
& \multicolumn{2}{|c|}{Whole sample} & \multicolumn{2}{|c|}{Women with regular cycles}\\
\hline
Variable & Mean & Standard deviation & Mean & Standard deviation \\
\hline
Mother's age & 29.69 & 6.18 & 29.46 & 6.05 \\
Mother's size (cm) & 164.58 & 9.30 & 164.47 & 9.09 \\
Mother's weight (kg) & 64.42 & 14.31 & 64.14 & 13.93 \\
Mother's Mass Corporal Index (MCI) & 24.51 & 14.73 & 24.16 & 12.69\\
Father's age & 30.69 & 6.88 &  30.79 & 6.93 \\
Father's size (cm) & 177.51 & 19.17 & 177.27 & 8.44\\
Father's weight (kg) & 81.13 & 40.30 & 81.96 & 45.34 \\
Father's MCI & 26.52 & 19.58 & 26.40 & 17.19\\
\hline
\end{tabular}
\caption{\textit{Characteristics of the parents.}}\label{tab:equationMC-comparaisonJV}
\end{center}
\end{table*}

\clearpage
\section{Tabulation of the CRL predicted by the IVF equation \eqref{eqFIVrob1:LCC=f(JV)}}

\begin{table*}[!h]
\begin{center}
\begin{tabular}{|cc|c|c||cc|c|c|}
\hline
 \multicolumn{2}{|c|}{FA} & CRL & SD of the  &  \multicolumn{2}{|c|}{FA} & CRL & SD of the \\
  days & weeks+days & mm &  CRL (mm) &   days & weeks+days & mm & CRL (mm)\\
  \hline
26 & 3 + 5 & 1.572894 & 3.047959 & 56 & 8 + 0 & 32.82732 & 2.308587\\
27 & 3 + 6 & 2.172470 & 2.913853 & 57 & 8 + 1 & 34.34187 & 2.431273\\
28 & 4 + 0 & 2.802544 & 2.781558 & 58 & 8 + 2 & 35.88692 & 2.557076\\
29 & 4 + 1 & 3.463118 & 2.651343 & 59 & 8 + 3 & 37.46247 & 2.685558\\
30 & 4 + 2 & 4.15419 & 2.523531 & 60 & 8 + 4 & 39.06852 & 2.816352\\
31 & 4 + 3 & 4.875762 & 2.398507 & 61 & 8 + 5 & 40.70507 & 2.949151\\
32 & 4 + 4 & 5.627834 & 2.276729 & 62 & 8 + 6 & 42.37212 & 3.083695\\
33 & 4 + 5 & 6.410404 & 2.158747 & 63 & 9 + 0 & 44.06967 & 3.219766\\
34 & 4 + 6 & 7.223474 & 2.045217 & 64 & 9 + 1 & 45.79771 & 3.357179\\
35 & 5 + 0 & 8.067043 & 1.936924 & 65 & 9 + 2 & 47.55626 & 3.495774\\
36 & 5 + 1 & 8.941112 & 1.834795 & 66 & 9 + 3 & 49.34530 & 3.635417\\
37 & 5 + 2 & 9.84568 & 1.739914 & 67 & 9 + 4 & 51.16485 & 3.775992\\
38 & 5 + 3 & 10.78075 & 1.653531 & 68 & 9 + 5 & 53.01489 & 3.917398\\
39 & 5 + 4 & 11.74631 & 1.577043 & 69 & 9 + 6 & 54.89544 & 4.059548\\
40 & 5 + 5 & 12.74238 & 1.511952 & 70 & 10+ 0 & 56.80648 & 4.202367\\
41 & 5 + 6 & 13.76894 & 1.459782 & 71 & 10 + 1 & 58.74802 & 4.345789\\
42 & 6 + 0 & 14.82601 & 1.421958 & 72 & 10 + 2 & 60.72006 & 4.489756\\
43 & 6 + 1 & 15.91357 & 1.399643 & 73 & 10 + 3 & 62.7226 & 4.634218\\
44 & 6 + 2 & 17.03163 & 1.393582 & 74 & 10 + 4 & 64.75564 & 4.779128\\
45 & 6 + 3 & 18.18019 & 1.403985 & 75 & 10 + 5 & 66.81918 & 4.924449\\
46 & 6 + 4 & 19.35925 & 1.430494 & 76 & 10 + 6 & 68.91322 & 5.070144\\
47 & 6 + 5 & 20.56881 & 1.472238 & 77 & 11 + 0 & 71.03775 & 5.216183\\
48 & 6 + 6 & 21.80887 & 1.527970 & 78 & 11 + 1 & 73.19279 & 5.362536\\
49 & 7 + 0 & 23.07943 & 1.596224 & 79 & 11 + 2 & 75.37833 & 5.50918\\
50 & 7 + 1 & 24.38049 & 1.675472 & 80 & 11 + 3 & 77.59436 & 5.656091\\
51 & 7 + 2 & 25.71204 & 1.764232 & 81 & 11 + 4 & 79.8409 & 5.803249\\
52 & 7 + 3 & 27.0741 & 1.861145 & 82 & 11 +5 & 82.11793 & 5.950636\\
53 & 7 + 4 & 28.46666 & 1.965003 & 83 & 11 + 6 & 84.42546 & 6.098235\\
54 & 7 + 5 & 29.88971 & 2.074765 & 84 & 12 + 0 & 86.76349 & 6.246032\\
55 & 7 + 6 & 31.34326 & 2.189542 & 85 & 12 + 1 & 89.13202 & 6.394012\\
  \hline
\end{tabular}
\caption{\textit{Predictions of the CRL as a function of FA for IVF data. We use Equations \eqref{eqFIVrob1:LCC=f(JV)} and the associated equation for the standard deviation (SD).}}
\end{center}
\end{table*}

\clearpage
\section{Tabulation of the FA predicted by the IVF equation \eqref{eq:JV-FIV}}

\begin{table*}[!h]
\begin{center}
\begin{tabular}{|c|c|c||c|c|c|}
\hline
 CRL & FA & SD of FA &  CRL & FA & SD of FA  \\
mm & days & days & mm & days & days \\
  \hline
1 & 23.92126 & 2.420171 & 43 & 62.06217 & 1.957944\\
2 & 26.43405 & 2.312322 & 44 & 62.64861 & 1.969268\\
3 & 28.39823 & 2.234829 & 45 & 63.23018 & 1.980882\\
4 & 30.07841 & 2.173452 & 46 & 63.80702 & 1.992769\\
5 & 31.57711 & 2.122639 & 47 & 64.37931 & 2.004914\\
6 & 32.9469 & 2.079527 & 48 & 64.94718 & 2.017301\\
7 & 34.21900 & 2.042402 & 49 & 65.51077 & 2.029915\\
8 & 35.41377 & 2.010139 & 50 & 66.07021 & 2.042745\\
9 & 36.54533 & 1.981947 & 51 & 66.62562 & 2.055776\\
10 & 37.62399 & 1.957245 & 52 & 67.17713 & 2.068998\\
11 & 38.65749 & 1.93559 & 53 & 67.72484 & 2.082398\\
12 & 39.65189 & 1.916631 & 54 & 68.26886 & 2.095966\\
13 & 40.61198 & 1.900084 & 55 & 68.8093 & 2.109691\\
14 & 41.54165 & 1.885717 & 56 & 69.34624 & 2.123565\\
15 & 42.44410 & 1.873333 & 57 & 69.87979 & 2.137577\\
16 & 43.32203 & 1.862764 & 58 & 70.41003 & 2.151719\\
17 & 44.1777 & 1.853865 & 59 & 70.93705 & 2.165983\\
18 & 45.01304 & 1.846509 & 60 & 71.46093 & 2.180362\\
19 & 45.82972 & 1.840584 & 61 & 71.98175 & 2.194847\\
20 & 46.62921 & 1.835989 & 62 & 72.49958 & 2.209431\\
21 & 47.41277 & 1.832634 & 63 & 73.0145 & 2.224109\\
22 & 48.18154 & 1.830438 & 64 & 73.52656 & 2.238874\\
23 & 48.9365 & 1.829326 & 65 & 74.03585 & 2.25372\\
24 & 49.67856 & 1.829231 & 66 & 74.54242 & 2.268641\\
25 & 50.4085 & 1.830088 & 67 & 75.04634 & 2.283632\\
26 & 51.12706 & 1.831841 & 68 & 75.54766 & 2.298688\\
27 & 51.83487 & 1.834435 & 69 & 76.04645 & 2.313803\\
28 & 52.53254 & 1.837822 & 70 & 76.54275 & 2.328975\\
29 & 53.2206 & 1.841953 & 71 & 77.03662 & 2.344197\\
30 & 53.89954 & 1.846786 & 72 & 77.52812 & 2.359466\\
31 & 54.56982 & 1.852281 & 73 & 78.01728 & 2.374779\\
32 & 55.23185 & 1.858398 & 74 & 78.50416 & 2.390131\\
33 & 55.88601 & 1.865103 & 75 & 78.9888 & 2.405518\\
34 & 56.53267 & 1.872362 & 76 & 79.47126 & 2.420939\\
35 & 57.17215 & 1.880142 & 77 & 79.95157 & 2.436388\\
36 & 57.80475 & 1.888415 & 78 & 80.42977 & 2.451864\\
37 & 58.43076 & 1.897152 & 79 & 80.9059 & 2.467364\\
38 & 59.05045 & 1.906327 & 80 & 81.38001 & 2.482884\\
39 & 59.66406 & 1.915915 & 81 & 81.85213 & 2.498423\\
40 & 60.27183 & 1.925892 & 82 & 82.3223 & 2.513977\\
41 & 60.87396 & 1.936236 & 83 & 82.79055 & 2.529545\\
42 & 61.47068 & 1.946927 & 84 & 83.25691 & 2.545124\\
  \hline
\end{tabular}
\caption{\textit{Predictions of the FA as a function of CRL by using Equations \eqref{eq:JV-FIV} and the associated equation for the standard deviation (SD).}}
\end{center}
\end{table*}

\clearpage
\section{Tabulation of the CRL predicted by the equation built on spontaneous pregnancies \eqref{eq:LCC-O2=f(JV)}}

\begin{table*}[!h]
\begin{center}
\begin{tabular}{|cc|c|c||cc|c|c|}
\hline
 \multicolumn{2}{|c|}{FA} & CRL & SD of the  &  \multicolumn{2}{|c|}{FA} & CRL & SD of the \\
  days & weeks+days & mm &  CRL (mm) &   days & weeks+days & mm & CRL (mm)\\
  \hline
26 & 3 + 5 & 1.160552 & 0.8733835 & 56 & 8 + 0 & 32.1681 & 5.582057\\
27 & 3 + 6 & 1.764087 & 1.524546 & 57 & 8 + 1 & 33.66139 & 5.620191\\
28 & 4 + 0 & 2.397282 & 1.961599 & 58 & 8 + 2 & 35.18434 & 5.654726\\
29 & 4 + 1 & 3.060135 & 2.309479 & 59 & 8 + 3 & 36.73695 & 5.685728\\
30 & 4 + 2 & 3.752647 & 2.604172 & 60 & 8 + 4 & 38.31922 & 5.713254\\
31 & 4 + 3 & 4.474817 & 2.862153 & 61 & 8 + 5 & 39.93114 & 5.737353\\
32 & 4 + 4 & 5.226645 & 3.092623 & 62 & 8 + 6 & 41.57273 & 5.75807\\
33 & 4 + 5 & 6.008133 & 3.301349 & 63 & 9 + 0 & 43.24397 & 5.77544\\
34 & 4 + 6 & 6.819278 & 3.492232 & 64 & 9 + 1 & 44.94488 & 5.789494\\
35 & 5 + 0 & 7.660083 & 3.668060 & 65 & 9 + 2 & 46.67544 & 5.800255\\
36 & 5 + 1 & 8.530546 & 3.830904 & 66 & 9 + 3 & 48.43566 & 5.807742\\
37 & 5 + 2 & 9.430667 & 3.982359 & 67 & 9 + 4 & 50.22553 & 5.811968\\
38 & 5 + 3 & 10.36045 & 4.123679 & 68 & 9 + 5 & 52.04507 & 5.81294\\
39 & 5 + 4 & 11.31989 & 4.255875 & 69 & 9 + 6 & 53.89426 & 5.810659\\
40 & 5 + 5 & 12.30898 & 4.379772 & 70 & 10+ 0 & 55.77312 & 5.805122\\
41 & 5 + 6 & 13.32774 & 4.496057 & 71 & 10+ 1 & 57.68163 & 5.796319\\
42 & 6 + 0 & 14.37615 & 4.605306 & 72 & 10+ 2 & 59.6198 & 5.784235\\
43 & 6 + 1 & 15.45422 & 4.708009 & 73 & 10 + 3 & 61.58763 & 5.76885\\
44 & 6 + 2 & 16.56196 & 4.804586 & 74 & 10 + 4 & 63.58512 & 5.750137\\
45 & 6 + 3 & 17.69935 & 4.895399 & 75 & 10 + 5 & 65.61226 & 5.728064\\
46 & 6 + 4 & 18.86639 & 4.980764 & 76 & 10 + 6 & 67.66907 & 5.702591\\
47 & 6 + 5 & 20.0631 & 5.060957 & 77 & 11 +0 & 69.75553 & 5.673673\\
48 & 6 + 6 & 21.28947 & 5.136219 & 78 & 11+ 1 & 71.87165 & 5.641257\\
49 & 7 + 0 & 22.54549 & 5.206765 & 79 & 11 + 2 & 74.01743 & 5.605281\\
50 & 7 + 1 & 23.83117 & 5.272784 & 80 & 11 + 3 & 76.19287 & 5.565678\\
51 & 7 + 2 & 25.14651 & 5.334444 & 81 & 11 + 4 & 78.39797 & 5.522369\\
52 & 7 + 3 & 26.49151 & 5.391894 & 82 & 11 + 5 & 80.63273 & 5.475266\\
53 & 7 + 4 & 27.86617 & 5.445268 & 83 & 11 + 6 & 82.89714 & 5.424271\\
54 & 7 + 5 & 29.27049 & 5.494685 & 84 & 12 + 0 & 85.19121 & 5.369271\\
55 & 7 + 6 & 30.70446 & 5.54025 & 85 & 12 +1 & 87.51494 & 5.310144\\
  \hline
\end{tabular}
\caption{\textit{Predictions of the CRL as a function of FA for spontaneous pregnancies. We use Equations \eqref{eq:LCC-O2=f(JV)} and the associated equation for the standard deviation (SD).}}
\end{center}
\end{table*}

\clearpage
\section{Tabulation of the FA predicted by the equation built on spontaneous pregnancies \eqref{eq:JV-HT}}

\begin{table*}[!h]
\begin{center}
\begin{tabular}{|c|c|c||c|c|c|}
\hline
 CRL & FA & SD of FA &  CRL & FA & SD of FA  \\
mm & days & days & mm & days & days \\
  \hline
1 & 25.29526 & 5.330148 & 43 & 62.79687 & 3.235434\\
2 & 27.887 & 5.156509 & 44 & 63.34921 & 3.212509\\
3 & 29.89792 & 5.024458 & 45 & 63.89639 & 3.190083\\
4 & 31.60820 & 4.914002 & 46 & 64.43858 & 3.168145\\
5 & 33.12633 & 4.817394 & 47 & 64.97593 & 3.146684\\
6 & 34.50799 & 4.730656 & 48 & 65.50861 & 3.125689\\
7 & 35.78624 & 4.651428 & 49 & 66.03676 & 3.105150\\
8 & 36.98260 & 4.578167 & 50 & 66.56052 & 3.085059\\
9 & 38.11206 & 4.509806 & 51 & 67.08001 & 3.065407\\
10 & 39.1855 & 4.445563 & 52 & 67.59537 & 3.046185\\
11 & 40.21114 & 4.384852 & 53 & 68.10671 & 3.027386\\
12 & 41.19538 & 4.327215 & 54 & 68.61414 & 3.009001\\
13 & 42.14330 & 4.272289 & 55 & 69.11778 & 2.991024\\
14 & 43.05901 & 4.219779 & 56 & 69.61772 & 2.973448\\
15 & 43.94591 & 4.169442 & 57 & 70.11407 & 2.956267\\
16 & 44.80685 & 4.121076 & 58 & 70.60691 & 2.939473\\
17 & 45.64421 & 4.074508 & 59 & 71.09635 & 2.923062\\
18 & 46.46006 & 4.029594 & 60 & 71.58246 & 2.907028\\
19 & 47.25616 & 3.986206 & 61 & 72.06532 & 2.891364\\
20 & 48.03405 & 3.944234 & 62 & 72.54503 & 2.876067\\
21 & 48.79508 & 3.903583 & 63 & 73.02165 & 2.861130\\
22 & 49.54045 & 3.864167 & 64 & 73.49525 & 2.846549\\
23 & 50.2712 & 3.825912 & 65 & 73.96591 & 2.832319\\
24 & 50.98829 & 3.78875 & 66 & 74.4337 & 2.818435\\
25 & 51.69256 & 3.752622 & 67 & 74.89868 & 2.804894\\
26 & 52.38477 & 3.717474 & 68 & 75.3609 & 2.791691\\
27 & 53.06561 & 3.683256 & 69 & 75.82045 & 2.778822\\
28 & 53.7357 & 3.649925 & 70 & 76.27736 & 2.766283\\
29 & 54.39563 & 3.617441 & 71 & 76.7317 & 2.754070\\
30 & 55.0459 & 3.585767 & 72 & 77.18352 & 2.742179\\
31 & 55.687 & 3.554868 & 73 & 77.63288 & 2.730607\\
32 & 56.31936 & 3.524715 & 74 & 78.07982 & 2.71935\\
33 & 56.94341 & 3.495280 & 75 & 78.52439 & 2.708405\\
34 & 57.5595 & 3.466535 & 76 & 78.96665 & 2.697768\\
35 & 58.16799 & 3.438456 & 77 & 79.40662 & 2.687436\\
36 & 58.7692 & 3.411022 & 78 & 79.84437 & 2.677406\\
37 & 59.36343 & 3.384211 & 79 & 80.27994 & 2.667675\\
38 & 59.95097 & 3.358004 & 80 & 80.71335 & 2.658239\\
39 & 60.53208 & 3.332383 & 81 & 81.14466 & 2.649096\\
40 & 61.107 & 3.307330 & 82 & 81.5739 & 2.640242\\
41 & 61.67596 & 3.282831 & 83 & 82.00112 & 2.631674\\
42 & 62.23918 & 3.258870 & 84 & 82.42634 & 2.623390\\
  \hline
\end{tabular}
\caption{\textit{Predictions of the FA as a function of CRL by using Equations \eqref{eq:JV-HT} and the associated equation for the standard deviation (SD).}}
\end{center}
\end{table*}

\end{document}